\documentclass[12pt]{elsarticle}
\usepackage{amsmath,amssymb}
\usepackage{graphicx}% Include figure files
\usepackage{caption}
\usepackage{color}% Include colors for document elements
\usepackage{dcolumn}% Align table columns on decimal point
\usepackage{bm}% bold math

\usepackage{amssymb}
\newcommand{\beq}{\begin{equation}}
\newcommand{\eeq}{\end{equation}}
\def\be{\begin{equation}} \def\ee{\end{equation}}
\def\ba{\begin{array}} \def\ea{\end{array}}
\def\bea{\begin{eqnarray}} \def\eea{\end{eqnarray}}

\journal{Physics Letters A}

\begin{document}

\begin{frontmatter}



\title{Configurational Entropy of Optical Bright Similariton in Tapered Graded-Index Waveguide}


\author[label1]{Pooja Thakur}\ead{poojaphy19@gmail.com}
\author[label2] {Marcelo Gleiser}\ead{mgleiser@dartmouth.edu}
\author[label3] {Anil Kumar}\ead{anilkumarphys@gmail.com}
 \author[label1] {Rama Gupta}\ead{Corresponding author:rama.gupta@gmail.com}
\address[label1]{Department of Physics, D. A. V. University,
Jalandhar-144 012, Punjab, India}
\address[label2]{Department of Physics and Astronomy, Darmouth
College, Hanover, New Hampshire 03755, USA}
\address[label3]{Department of Physics, JC DAV College (Panjab University), Dasuya-144 205, Punjab, India}

\begin{abstract}

\noindent Configurational entropy (CE) consists of a family of entropic measures of
information used to describe the shape complexity of spatially-localized functions with
respect to a set of parameters. We obtain the Differential Configurational Entropy (DCE) for similariton
waves traveling in tapered graded-index optical waveguides
modeled by a generalized nonlinear Schr\"odinger equation.
It is found that for similariton's widths lying within a certain
range, DCE attains minimum saturation values as the nonlinear wave
evolves along the effective propagation variable $\zeta(t)$. In particular, saturation is achieved earlier for lower values of the width, which we show correspond to global minima of the DCE. Such low entropic values
lead to minimum dispersion of
momentum modes as the similariton waves propagate along tapered graded-index
waveguides, and should be of importance in guiding their design.
\end{abstract}


\end{frontmatter}


\section{Introduction}
\label{1} \noindent Nonlinear phenomena appear in all areas of
science, from physics and chemistry to the life sciences and
engineering \cite{rajaraman1, dodd2}. In most applications, nonlinear systems are
modelled by partial
differential equations called nonlinear evolution equations
(NLEEs). Of much interest are solutions in the shape of
spatially-localized excitations. Many such excitations are
non-dispersive and non-dissipative waves, that is, configurations that are
weakly or non-radiative and that maintain their shape as they propagate to significantly
long distances \cite{russel3}. Due to this special feature, these configurations
are known as solitons or solitary waves.
Soliton-like solutions appear in various research areas, including
hydrodynamics, plasma physics, nonlinear optics, condensed matter
physics, optical communications, nuclear physics, and astrophysics
\cite{miki4, dudarev5, christ6, schieff7, sievers8, trombe9}.

Different strategies are used to investigate the dynamical
features of soliton-like solutions from NLEEs, including exact
analytical methods \cite{painleve10}, numerical techniques
\cite{lakshman11}, variational analysis, and other tools
\cite{ansari12}. A new emerging area is the study of the
configurational entropy (CE) of these nonlinear excitations, a
measure of spatial complexity of spatially-localized systems
proposed by Gleiser and Stamatopoulos, originally applied to the
study of kinks and bounces in relativistic field theory
\cite{Nikitas14}. As described in detail in Ref.
\cite{Stephens23}, the original CE has been expanded to include a
family of related entropic measures, Configuration Information
Measures (CIMs), adapted for different applications. In the
current case, for a continuum field theory, the proper measure is
known as Differential Configurational Entropy (DCE), as defined
below. It gives a quantitative measure of the shape complexity of
a function or, in the case of most physical applications, of a
given field configuration.

Inspired by Shannon information entropy \cite{shannon13}, the DCE
is obtained from the Fourier transform of a spatially-localized
configuration ({\it e.g.} a solution of a NLEE) or its related
energy density. There is an intimate link between information and
dynamics, where the entropic measure plays a prominent role in
signaling the emergence of nonlinear structures and in
highlighting their stability properties \cite{Nikitas14,
gleiser15}. Configurational Entropy has since been applied to a
variety of fields, including neutron and boson stars
\cite{mgleiser16}, spontaneous symmetry breaking \cite{gleiser15},
glueballs \cite{bernardini17}, the stability of Q-balls
\cite{gleiserm18}, anti-de Sitter black holes \cite{braga19,
lee20}, stability of graviton Bose-Einstein condensate in the
brane-world \cite{roberto21}, the energy-energy correlation in
$e^{+}e^{-}$ into hadrons \cite{karapetyan22}, oscillon lifetimes
in scalar field theories \cite{Stephens23}, instantons and vacuum
decay in arbitrary spatial dimensions \cite{Sowinski24}, dynamical
tachyonic holographic Ads/QCD models \cite{Barbosa25}, standard
model cosmology for the homogeneous Friedmann-Robertson-Walker
universe \cite{abernardini26} and in inflationary cosmology
\cite{gleisergraham27}, and in the study of dissociation of heavy
vector mesons in a thermal medium \cite{Braga28}. Apart from the
above applications in high-energy physics and cosmology, DCE has
been calculated for several nonlinear scalar field models
featuring solutions with spatially-localized energy, including
bounces in one spatial dimension and critical bubbles in three
spatial dimensions \cite{Nikitas14}, for the determination of
critical points in second order phase transitions
\cite{Sowinski29}, solitons in supersymmetric theories
\cite{correa30}, and for Korteweg-de Vries solitons in quark-gluon
plasmas \cite{goncalves31}.

In the present work, we apply the DCE measure
to solitary waves propagating along tapered graded-index nonlinear
waveguides. The graded potential is widely used
in optical fiber communications \cite{arrue32}, computer networks
\cite{Sengupta33}, long-distance telecommunications \cite{Qiao34},
and sensory receptor cells \cite{Ping35}. It solves the
problem of modal dispersion to a considerable extent
\cite{Sengupta33}. In particular, we will investigate the propagation of waves in
tapered graded-index waveguides modeled by a
generalized nonlinear Schr\"odinger equation (GNLSE)
\cite{agrawal36, ponomarenko37, rama38} and obtain the
DCE for an optical similariton, a solitonic solution of the GNLSE
with variable coefficient.

The paper is organized as follows: In section $2$, we briefly
review DCE. (For a more detailed discussion see Ref.
\cite{Stephens23}.) In section $3$, the similariton solution of
GNLSE is obtained invoking the similarity transformation, as well
as the saturation minimum-value DCE as a function of width as the
similariton evolves in time. We also find that the similariton's
DCE has global minima as a function of width. These time-dependent
values determine the optimal value of the width for the
configurational stability of similaritons propagating through
tapered graded-index optical waveguides. Finally, we summarize our
results in section $4$.

\section {Differential Configurational Entropy for bright similaritons in tapered
graded-index waveguide} \label{2} \noindent The Differential
Configuration Entropy of a square-integrable, bounded mathematical
function is constructed from its Fourier modes \cite{Nikitas14,
Stephens23}. In physical applications, it is often the case that
the
 square-integrable continuous function is the energy density $\rho(x)$
defined on $R^{d}$ and with Fourier transform, \beq
\label{fourier}  F(k)=\int\exp[-i x\cdot k]\rho(x) d^{d}x. \eeq
Henceforth, we consider only $d=1$. The corresponding modal
fraction, which measures the relative weight of each mode $k$ is
defined as \cite{Nikitas14} \beq \label{modal fraction}
 f(k)=\frac{|F(k)|^{2}}{\int|F(k)|^{2}dk}.  \eeq
To ensure positivity of DCE, the modal fraction is normalized by
the mode carrying
maximum weight $f_{\rm max}(k),$
 \beq \label{nor}
\tilde{f}(k)=\frac{f(k)}{f_{\rm max}(k)}.\eeq
 The mathematical expression of
DCE is defined as $S_{c}[\tilde{f}]$ is \cite{Nikitas14}

\beq \label{configuration} S_{c}[\tilde{f}
]=-\int_{-\infty}^{\infty}\tilde{f}(k)\ln\tilde{f}(k)dk, \eeq

\noindent which represents an absolute limit on the best lossless
compression of any communication \cite{shannon39}.
For periodic functions, one would use the Fourier series of the function
$\rho(x)$ to define DCE. While other possible functional
transforms could in principle be used to obtain the DCE, the clear
physical interpretation of Fourier transform relating increased spatial localization to broader
momentum-mode distribution, makes it the most efficient to define DCE,
as the many applications referenced above have shown.

\section {Differential Configurational Entropy for bright similaritons in tapered
graded-index waveguide}\label{3}\subsection {\bf Model equation
and similariton solution} \noindent The beam propagation in
tapered graded-index nonlinear fiber waveguide is modelled by the
GNLSE \cite{agrawal36, ponomarenko37, rama38}. The GNLSE in
$(1+1)$-dimensional form is given as

\beq\label{gnlse} i\frac{\partial \Psi}{\partial
t}+\frac{1}{2}a^{2}(t)\frac{\partial^{2}\Psi}{\partial
x^{2}}+\frac{1}{2}M(t)x\Psi-\frac{i}{2}g(t)\Psi+2\gamma(t)\Psi+\mu|\Psi|^{2}\Psi=0.
\eeq

\noindent Here $a(t), M(t)$, and $\gamma(t)$ are the time-dependent dispersion
coefficient, tapered potential, and external potential,
respectively, while
$x$ is the transverse direction and $\Psi$ is the wave function.
Further, $g(t)$ is a dimensionless net energy gain (if $g>0$) or loss (if $g<0$) in the system, and $\mu$ is the coefficient
of nonlinearity.

Introducing the similarity variable $\chi$,
 \beq
\label{space} \chi(x,t)= \frac{[x-x_{c}(t)]}{\alpha(t)},
 \eeq
the self-similar optical similariton solution of Eq.
(\ref{gnlse}) can be obtained by transforming it into a standard
nonlinear Schr\"odinger equation (NLSE) by using gauge and
similarity transformations \cite{agrawal36, ponomarenko37, rama38}
\beq\label{similarity} \Psi(x,t)=B(t) \Phi\left[\chi(x,t),
\zeta(t)\right] \exp[{i\varphi(x,t)}], \eeq where $B(t),
\alpha(t)$, and $x_{c}(t)$ are the dimensionless amplitude, width,
and guiding-center coordinate of the beam, respectively.

Assuming a linear ansatz for the global phase,
\beq \label{global}
 \varphi(x,t)=[p_{1}(t)x+p_{2}(t)],  \eeq
 and substituting Eqs. (\ref{similarity}) and (\ref{global}) into Eq.
(\ref{gnlse}), we obtain a set of first-order differential equations for the
parameters of the transformation and for the transformed field
$\Phi$, which satisfies the standard NLSE \cite{zakhrov40}
\beq
\label{nlse}
i\frac{\partial\Phi}{\partial\zeta}+\frac{1}{2}\frac{\partial^{2}\Phi}{\partial\chi^{2}}+\mu|\Phi|^{2}\Phi=0.
\eeq
Here, the effective propagation variable $\zeta(t)$ and
guiding-center position $x_{c}(t)$ are given by
\beq\label{effdistance}
 \zeta(t)=\zeta_{0}+\int_{0}^{t}B^{2}(\tau)d\tau,
\eeq  \beq \label{guiding}
x_{c}(t)=\alpha(t)\left[\int_{0}^{t}\frac{a^{2}(\tau)p_{1}(\tau)d\tau}{\alpha
(\tau)}+x_{0}\right],  \eeq where $\zeta_{0}$ and $x_{0}$ are
integration constants which we fix as $\zeta_{0}=1,
x_{0}=1$. The remaining nonlinear partial differential equations for the parameters are
$$B(t)=\frac{a(t)}{\alpha},$$
$$g(t)= 2\frac{\dot{B}(t)}{B(t)},$$
$$M(t)=2\dot{p_{1}}(t),$$
$$\gamma(t)=\frac{1}{2}\dot{p_{1}}(t)+a^{2}(t)p_{1}^{2}(t).$$

\noindent Eq. (\ref{nlse}) is a standard NLSE, admitting bright
\cite{zakhrov40} and dark \cite{Zakhrav41} soliton solutions, as
well as rogue waves and breathers \cite{akhmediev42}. Using these,
and width $\alpha(t)$ constant, one can obtain solutions of Eq.
(\ref{gnlse}) implementing the self-similar transformation of Eq.
(\ref{similarity}). In particular, the bright similariton solution
of Eq. (\ref{nlse}) is given as \beq\label{soliton solution}
\Psi(x, \zeta(t))= a\left(\sqrt{\frac{\mu^{3}}{2}}\hbox{sech}\left
(\mu^{2}\frac{(\sqrt{2}\left(\frac{x-x_{c}}{\alpha}\right)-\chi_{0})}{2}+2\xi\zeta
\mu^{2}\right)\exp(r_{1})\right)r_{0} , \eeq where
$$r_{0}=\exp\left(i(p_{1}\cdot x+p_{2})\right),$$

$$r_{1}=-4i\left(\xi^{2}-\left(\frac{\mu^{2}}{4}\right)^{2}\right)\zeta-2i\xi\sqrt{2}\left(\frac{x-x_{c}}{\alpha}\right)+i\epsilon.$$

\noindent In Eq. (\ref{soliton solution}), the dispersion coefficient
$a(t)$ is the amplitude of the similariton, which, in general, could have
space and time dependence. Here, we have chosen it to be a periodic
function of $t$,
$$ a(t)=\frac{1+\cos^{2}(t)}{\alpha} .$$

\label{4}\subsection {\bf Calculation of Differential Configurational Entropy}
\noindent The energy density corresponding to the
solution (Eq. (\ref{soliton solution})) can be written as

\beq \label{posdensity} \rho(x,
\zeta(t))=\frac{\mu^{2}\hbox{sech}\left[2\mu^{2}\zeta\xi+\frac{1}{2}\mu^{2}\left(\frac{\sqrt{2}(x-x_{c})}{\alpha}-\chi_{0}\right)\right]^{2}}{2\sqrt{2}\alpha}.
 \eeq
 The square of the Fourier transform of the energy density $F^{2}(k)$, using
Eq.(\ref{fourier}), is given as

\beq \label{momdensity}  F^{2}(k,
\zeta(t))=\frac{r_{2}\left(\frac{-i r_{3}(-i
k\alpha+\sqrt{2}\mu^{2})}{r_{3}+r_{4}}+ k\alpha
r_{5}\right)\left(\frac{i r_{3}(i
k\alpha+\sqrt{2}\mu^{2})}{r_{3}+r_{4}}+ k\alpha r_{7}\right)}
{\pi(k^2\alpha^2 +2\mu^2)},
\eeq where

$$r_{2}=\frac{\exp\left(-2\mu^{2}(\sqrt{2}x_{c}+\alpha(-4\zeta\xi+\chi_{0}))\right)}{\alpha},$$

$$r_{3}=\frac{\exp\left(\mu^{2}(\sqrt{2}x_{c}+\alpha\chi_{0})\right)}{\alpha},$$

$$r_{4}=\exp(4\zeta\xi\mu^{2}),$$

 $$r_{5}=2F1\left[1,1-\frac{i
k\alpha}{\sqrt{2}\mu^{2}},2-\frac{i k\alpha}{\sqrt{2}\mu^{2}},
r_{6}\right],$$

$$r_{6}=\frac{\exp\left(-\mu^{2}(\sqrt{2}x_{c}+\alpha(-4\zeta\xi+\chi_{0}))\right)}{\alpha},$$

$$r_{7}=2F1\left[1,1+\frac{i
k\alpha}{\sqrt{2}\mu^{2}},2+\frac{i k\alpha}{\sqrt{2}\mu^{2}},
r_{6}\right],$$

\begin{figure}
\begin{center}
\vspace{-0.2cm}
\begin{tabular}{ccc}
\includegraphics[width=2.5in]{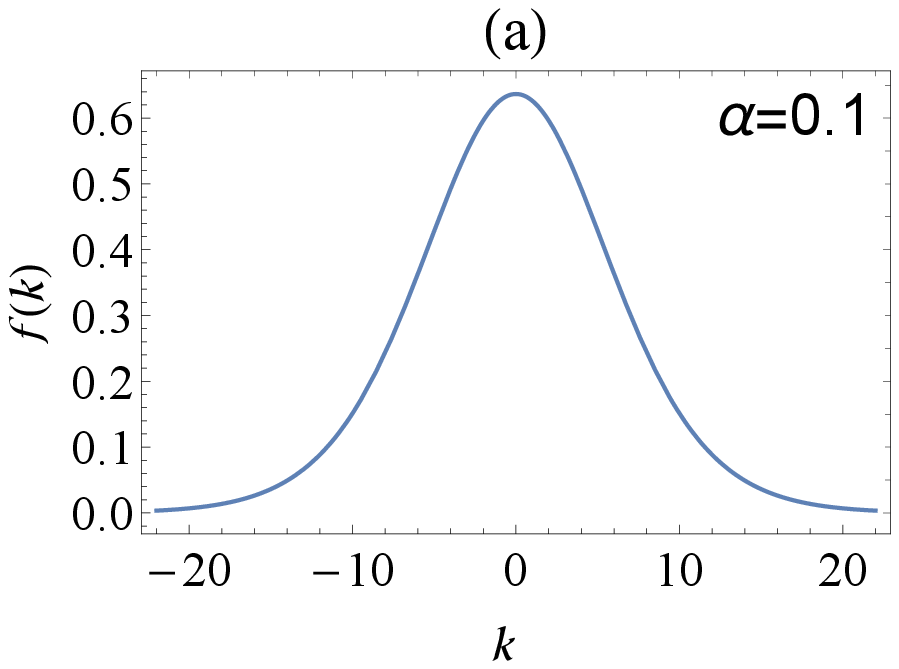}
\includegraphics[width=2.5in]{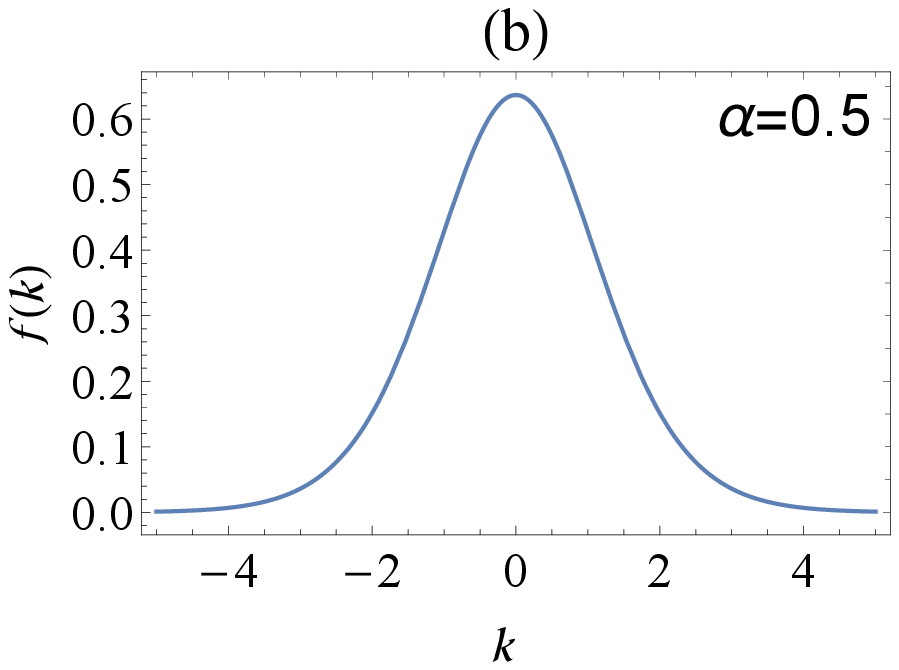}
\vspace{0.2cm}  \\
\includegraphics[width=2.5in]{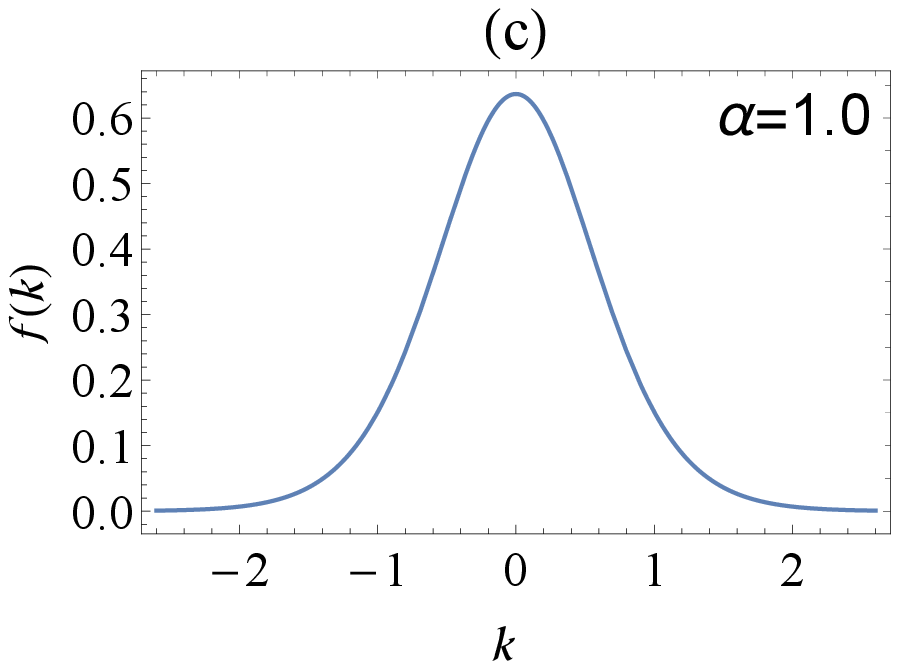}
\includegraphics[width=2.5in]{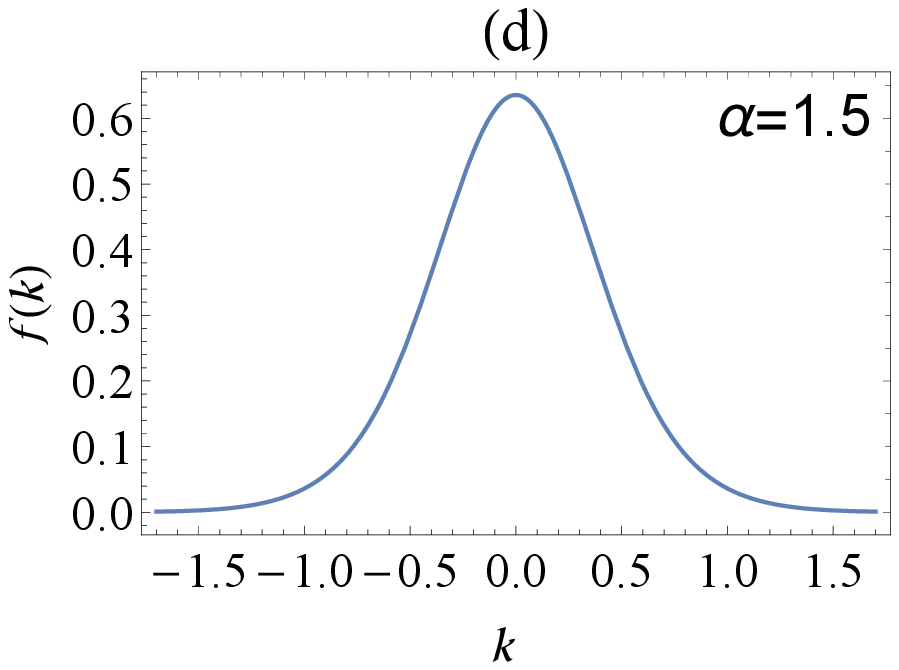}
\vspace{0.2cm}  \\
\end{tabular}
\caption{\label{fig1} The modal fractions for bright similaritons
with $\alpha=0.1, 0.5, 1.0, 1.5$ showing a maximum at $k=0$.}
\end{center}
\end{figure}

\noindent and $2F1$ is the Hypergeometric function.

From a detailed investigation of the dependence of  $F^{2}(k)$
(Eq. (\ref{momdensity})) on its parameters,  we observed that DCE
depends most strongly on the width $\alpha$ of the similariton. To
study this dependence, we fixed the values of the other parameters
as $x_{0}=1, \chi_{0}=0.3, \zeta_{0}=1, \epsilon=1, \mu=1, \xi=1$.
In Fig. \ref{fig1}, we plot the modal fraction of the bright
similariton solution for several values of $\alpha$, showing that
all have a clear bell-shaped curve with a maximum at $k=0$.

\begin{figure}[ht]
\begin{center}
\begin{tabular}{ccc}
\vspace{-0.2cm}
\includegraphics[width=2.7in]{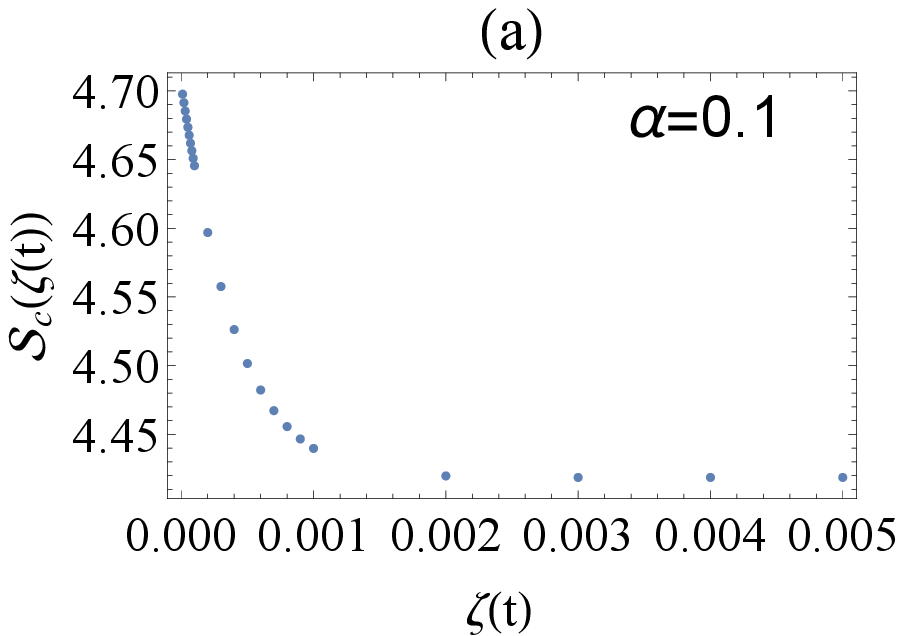}
\vspace{0.2cm}
\includegraphics[width=2.6in]{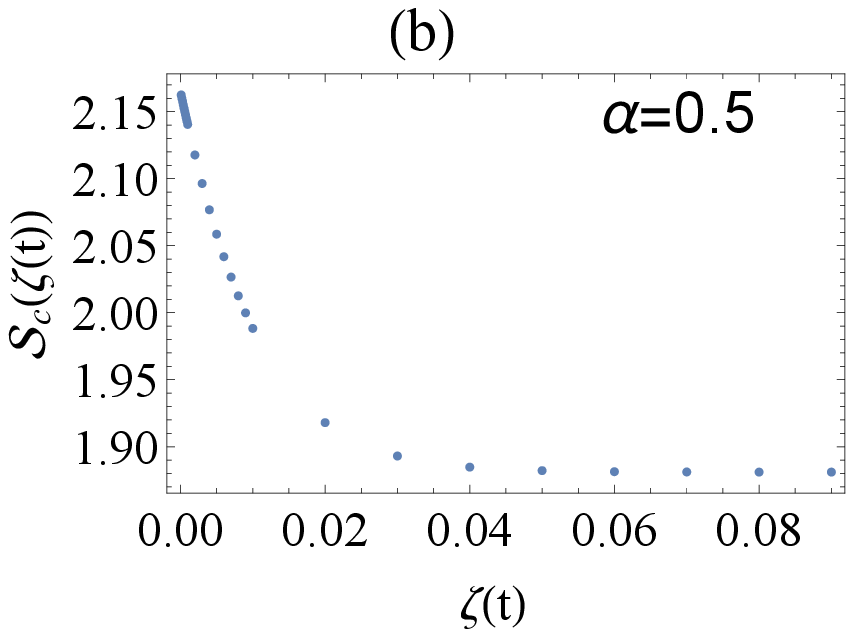}
\vspace{0.2cm} \\
\includegraphics[width=2.7in]{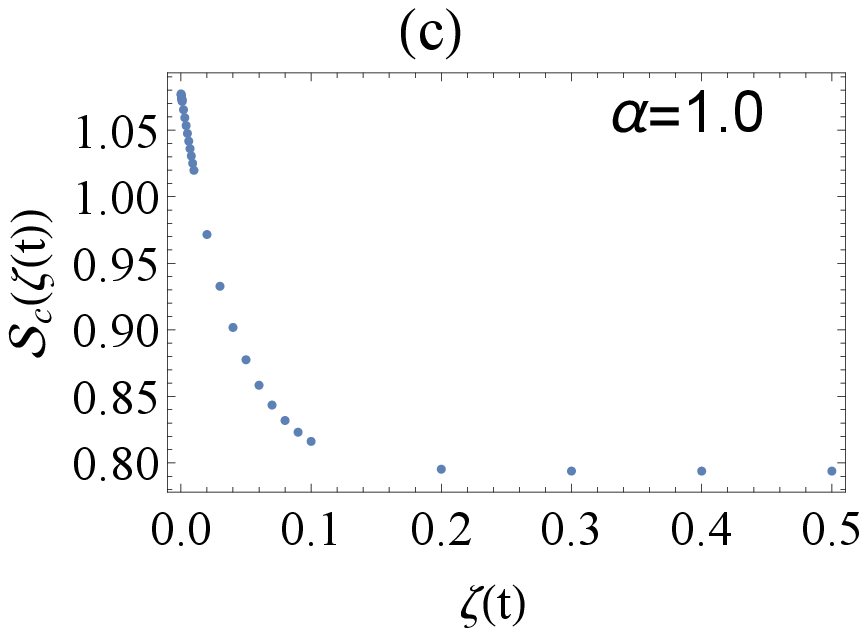}
\vspace{0.2cm}
\includegraphics[width=2.6in]{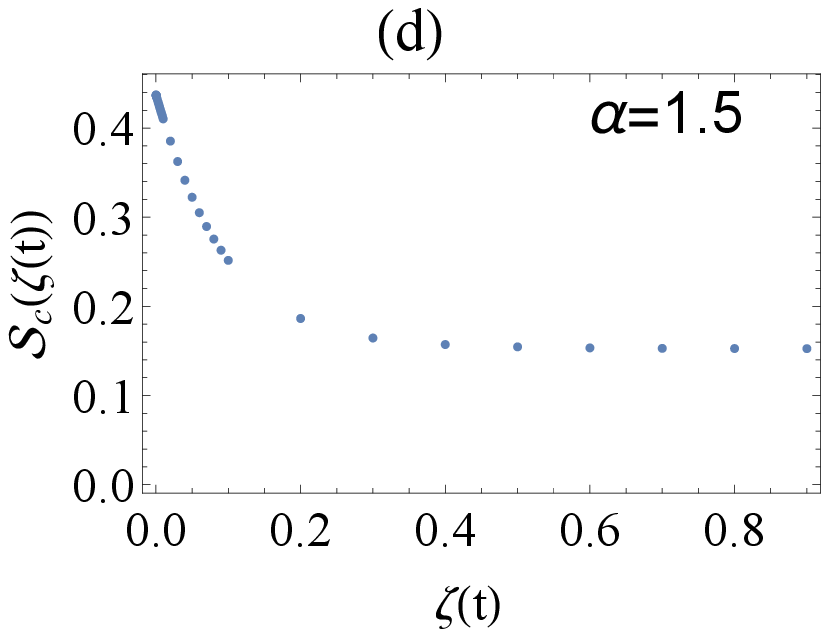}
\vspace{-0.2cm}
\end{tabular}
\caption{\label{fig2} Differential Configurational entropy of
bright optical similariton in the tapered graded-index waveguide
propagating in the $\zeta(t)$ time at similariton width (a)
$\alpha=0.1$, (b) $\alpha=0.5$, (c) $\alpha=1.0$, (d)
$\alpha=1.5$.}
\end{center}
\end{figure}

\begin{figure}[ht]
\begin{center}
\begin{tabular}{ccc}
\vspace{-0.2cm}
\includegraphics[width=2.5in]{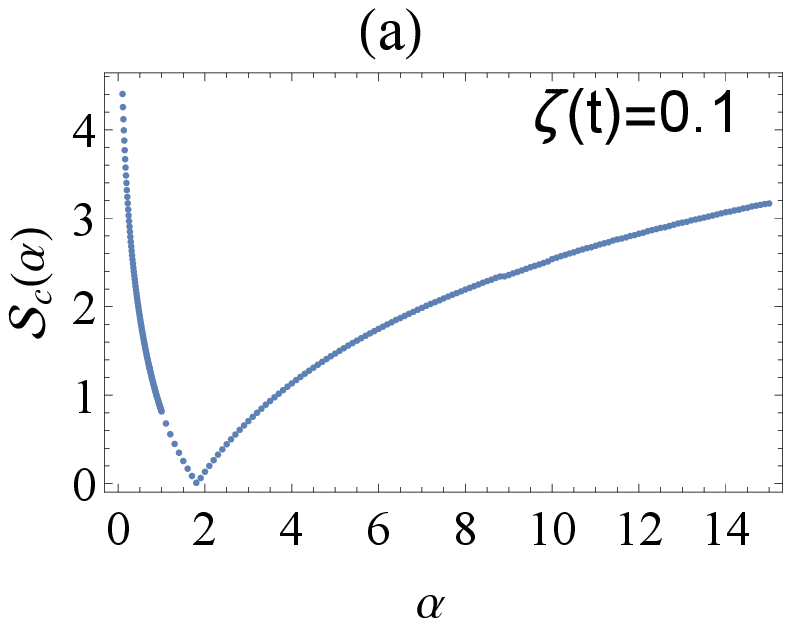}
\vspace{0.2cm}
\includegraphics[width=2.5in]{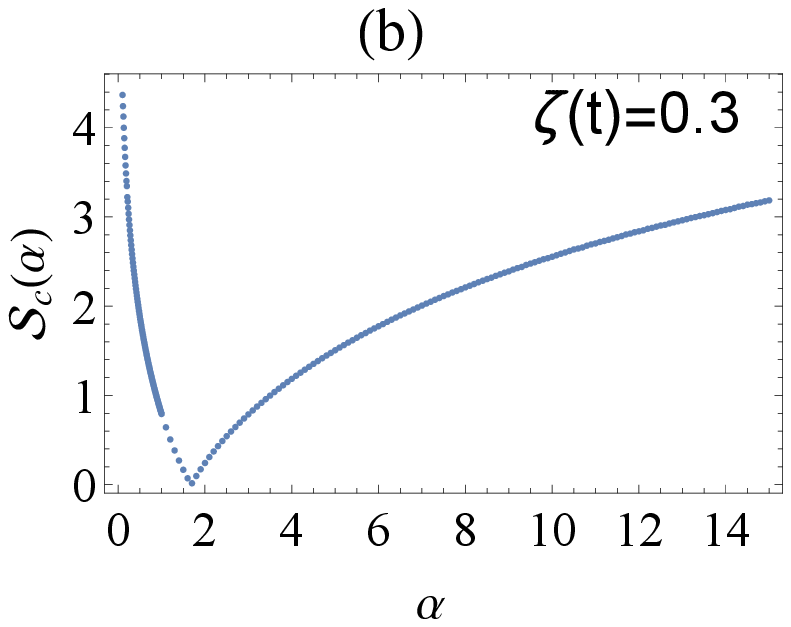}
\vspace{0.2cm} \\
\includegraphics[width=2.5in]{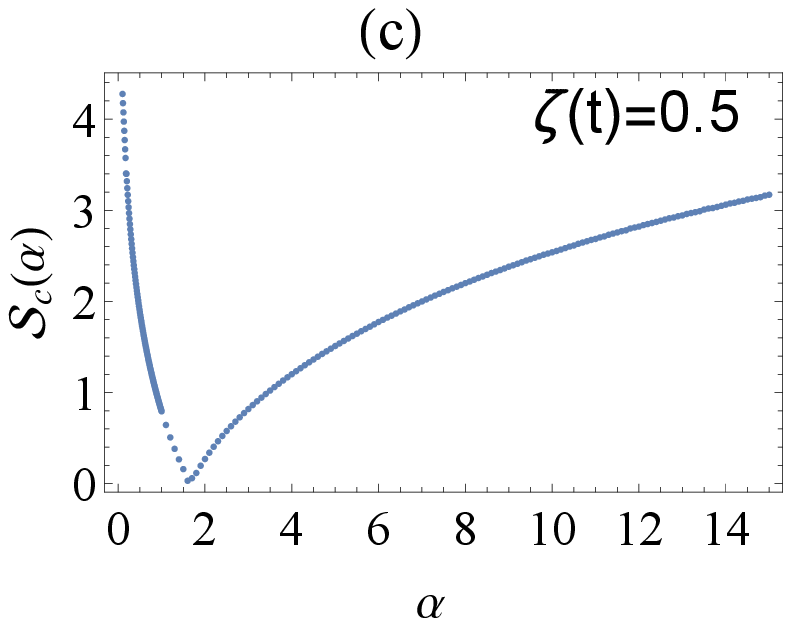}
\vspace{0.2cm}\\
\end{tabular}
\caption{\label{fig3} Differential configurational entropy ($S_{c}(\alpha)$) as
a function of the similariton width $\alpha$. The minimum value of
DCE ($S_{c}(\alpha)$) occurs at (a) $\alpha=1.8$, (b) $\alpha=1.7$,
(c) $\alpha=1.6$.}
\end{center}
\end{figure}

The evolution of DCE  for the bright similariton (given by Eq.
(\ref{soliton solution})) as a function of the time parameter
$\zeta(t)$ was obtained using Eq. (\ref{configuration}) for
several values of $\alpha$. As shown in Fig. \ref{fig2}, DCE
saturates in all cases at a minimum value sensitive to the value
of $\alpha$. For $\alpha=0.1, 0.5, 1.0, 1.5$ the saturation occurs
at $\zeta(t)\simeq 0.001, 0.02, 0.1, 0.2$, respectively: the
smaller the width $\alpha$, the earlier the similariton reaches
the saturation value. The same behavior has been observed for all
$\alpha$ belonging to $[0.1, 1.5]$.

Alternatively, one may examine the dependence of
$S_c^{\alpha}(\zeta(t))$ on the width parameter $\alpha$ itself at
different fixed times in the evolution of the bright similariton.
The results are shown in Fig. \ref{fig3} for three snapshots of
$\zeta(t)=0.1, 0.3, 0.5$. Overall, we find that
$S_c^{\alpha}(\zeta(t))$ is weakly dependent on $\zeta(t)$, with a
global minimum at $\alpha\simeq 1.8, 1.7, 1.6$, respectively.
Furthermore, comparing Figs. \ref{fig2} and Fig. \ref{fig3} we
note that the saturation values for different $\alpha$ in Fig.
\ref{fig2} correspond to the values of $S_c^{\alpha}(\zeta(t))$ in
Fig. \ref{fig3}. This can be clearly seen comparing Figs.
\ref{fig2}(d) and \ref{fig3}(c), since the values of $\alpha$ for
both are near the global minimum of $S_c^{\alpha}(\zeta(t))$. As
discussed in the literature (see {\it e.g.} \cite{mgleiser16,
bernardini17, gleiserm18, Stephens23, Barbosa25, Braga28}), a
minimum of DCE signals the most stable configuration with respect
to a given parameter, in this case the similariton width $\alpha$.
These values denote the best range for the similariton's width to
ensure its propagation through the tapered graded-index waveguide
with optimal compression of information.

\section {Conclusion} \label{5}
\noindent The tapered graded-index waveguide finds applications in
optical fiber communications \cite{arrue32}, computer networks
\cite{Sengupta33}, long-distance telecommunications \cite{Qiao34},
and sensory receptor cells \cite{Ping35, Mikko43}. In this work,
we have computed the differential configurational entropy
($S_{c}^{\alpha}(\zeta(t)$) for optical bright similaritons
propagating along a tapered graded-index waveguide. We found that
$S_{c}^{\alpha}(\zeta(t)$) has a weakly time-dependent global
minimum for a narrow range of width $\alpha$. We can thus see that
this formalism helps one obtain the optimal width of the
similariton waves for which dispersion is minimized and the
spatial shape is most compressed into its momentum modes. This
optimal compression ensures the propagation of minimally-entropic
similariton waves through the tapered graded-index waveguide. In
future work, we may investigate the dependence of our results on
different values of the other similariton parameters and expand
our method to two and three spatial dimensions, computing the
related  differential configurational entropy (DCE)
\cite{Stephens23}. We hope this work will aid in the design of
more efficient tapered graded-index waveguides with optimized
widths.

\section {Acknowledgment} \label{6}
\noindent The financial support from Department of Science and
Technology, New Delhi through Women Scientist Scheme-A project
(Ref. No. SR/ WOS- A/ PM-109/2017 G) is gratefully acknowledged by
Pooja Thakur.





\end{document}